\documentclass[10pt,conference]{IEEEtran}
\usepackage{cite}
\usepackage{amsmath,amssymb,amsfonts}
\usepackage{algorithmic}
\usepackage{graphicx}
\usepackage{textcomp}
\usepackage{xcolor}
\usepackage[hyphens]{url}
\usepackage{fancyhdr}
\usepackage{hyperref}

\definecolor{olivegreen}{rgb}{0, 0.6, 0}
\definecolor{redorange}{HTML}{FF5349}
\definecolor{blue(ncs)}{rgb}{0.0, 0.53, 0.74}
\definecolor{navy}{HTML}{273BE2}
\definecolor{black}{HTML}{000000}
\definecolor{white}{HTML}{ffffff}

\usepackage{xspace}

\usepackage{graphicx}

\usepackage[font=scriptsize]{subcaption}

\usepackage{multicol}
\usepackage{multirow}
\usepackage{makecell}
\usepackage{booktabs}

\usepackage[noabbrev, capitalize]{cleveref}

\usepackage{siunitx}

\usepackage{comment}

\newcommand{\thiswork}{Smart-Infinity\xspace}

\newcommand{\rev}[1]{{#1}}

\newcommand{\smartupdate}{SmartUpdate\xspace}

\newcommand{\SmartUpdate}{SmartUpdate\xspace}

\newcommand{\SmartTopK}{SmartComp\xspace}

\usepackage{tikz}
\newcommand*\circled[1]{\tikz[baseline=(char.base)]{
            \node[shape=circle,draw,inner sep=0.4pt] (char) {#1};}}

\newcommand*\bcircled[1]{\tikz[baseline=(char.base)]{
            \node[shape=circle,draw,inner sep=0.4pt, fill=black, text=white] (char) {#1};}}

\newcommand{\hpcayear}{2024}

\newcommand{\hpcasubmissionnumber}{377}
\title{
\thiswork: Fast Large Language Model Training using Near-Storage Processing on a Real System
}

\def\hpcacameraready{} 

\newcommand\hpcaauthors{
        Hongsun Jang\IEEEauthorrefmark{2},
        Jaeyong Song\IEEEauthorrefmark{2},
        Jaewon Jung\IEEEauthorrefmark{2},
        Jaeyoung Park\IEEEauthorrefmark{3}\textsuperscript{,}\IEEEauthorrefmark{1},
        Youngsok Kim\IEEEauthorrefmark{4},
        Jinho Lee\IEEEauthorrefmark{2}\textsuperscript{,}\IEEEauthorrefmark{5}
    
}

\newcommand\hpcaaffiliation{
\IEEEauthorrefmark{2}\textit{Department of Electrical and Computer Engineering, Seoul National University}\\%
\IEEEauthorrefmark{3}\textit{Department of Electrical and Computer Engineering, University of Texas at Austin}\\%
\IEEEauthorrefmark{4}\textit{Department of Computer Science, Yonsei University}
}

\newcommand\hpcaemail{
    \{hongsun.jang, jaeyong.song, jungjaewon\}@snu.ac.kr,
    jaeyoung@utexas.edu,
    youngsok@yonsei.ac.kr,  leejinho@snu.ac.kr
}

\author{
  \ifdefined\hpcacameraready
    \IEEEauthorblockN{\hpcaauthors{}}
      \IEEEauthorblockA{
        \hpcaaffiliation{} \\
        \hpcaemail{}
      }
  \else
    \IEEEauthorblockN{\normalsize{HPCA \hpcayear{} Submission
      \textbf{\#\hpcasubmissionnumber{}}} \\
      \IEEEauthorblockA{
        Confidential Draft \\
        Do NOT Distribute!!
      }
    }
  \fi 
}

\fancypagestyle{camerareadyfirstpage}{%
  \fancyhead{}
  
  \fancyhead[C]{
    \ifdefined\aeopen
    \parbox[][12mm][t]{13.5cm}{\hpcayear{} IEEE International Symposium on High-Performance Computer Architecture (HPCA)}    
    \else
      \ifdefined\aereviewed
      \parbox[][12mm][t]{13.5cm}{\hpcayear{} IEEE International Symposium on High-Performance Computer Architecture (HPCA)}
      \else
      \ifdefined\aereproduced
      \parbox[][12mm][t]{13.5cm}{\hpcayear{} IEEE International Symposium on High-Performance Computer Architecture (HPCA)}
      \else
      \parbox[][0mm][t]{13.5cm}{\hpcayear{} IEEE International Symposium on High-Performance Computer Architecture (HPCA)}
    \fi 
    \fi 
    \fi 
    \ifdefined\aeopen 
      \includegraphics[width=12mm,height=12mm]{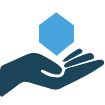}
    \fi 
    \ifdefined\aereviewed
      \includegraphics[width=12mm,height=12mm]{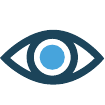}
    \fi 
    \ifdefined\aereproduced
      \includegraphics[width=12mm,height=12mm]{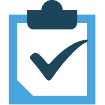}
    \fi
  }
  \fancyfoot[C]{}
}
\begin{document}
\maketitle

\ifdefined\hpcacameraready 
  \thispagestyle{camerareadyfirstpage}
  \pagestyle{empty}
\else
  \thispagestyle{plain}
  \pagestyle{plain}
\fi

\newcommand{\hpcaheight}{0mm}
\ifdefined\eaopen
\renewcommand{\hpcaheight}{12mm}
\fi



\let\svthefootnote\thefootnote

\newcommand\freefootnote[1]{%
  \let\thefootnote\relax%
  \footnotetext[0]{#1}%
  \let\thefootnote\svthefootnote%
}

\freefootnote{
    \IEEEauthorrefmark{1}Work performed while at Yonsei University. \IEEEauthorrefmark{5}Corresponding author.
}


\begin{abstract}
The recent huge advance of Large Language Models (LLMs) is mainly driven by the increase in the number of parameters. 
This has led to substantial memory capacity requirements, necessitating the use of dozens of GPUs just to meet the capacity.
One popular solution to this is storage-offloaded training, which uses host memory and storage as an extended memory hierarchy.
However, this obviously comes at the cost of storage bandwidth bottleneck because storage devices have orders of magnitude lower bandwidth compared to that of GPU device memories.

Our work, \thiswork, addresses the storage bandwidth bottleneck of storage-offloaded LLM training using near-storage processing devices on a real system. 
The main component of \thiswork is \smartupdate, which performs parameter updates on custom near-storage accelerators. 
We identify that moving parameter updates to the storage side removes most of the storage traffic.
In addition, we propose an efficient data transfer handler structure to address the system integration issues for \thiswork. 
The handler allows overlapping data transfers with fixed memory consumption by reusing the device buffer.
Lastly, we propose accelerator-assisted gradient compression/decompression to enhance the scalability of \thiswork. 
When scaling to multiple near-storage processing devices, the write traffic on the shared channel becomes the bottleneck.
To alleviate this, we compress the gradients on the GPU and decompress them on the accelerators. 
It provides further acceleration from reduced traffic.
As a result, \thiswork achieves a significant speedup compared to the baseline.
Notably, \thiswork is a ready-to-use approach that is fully integrated into PyTorch on a real system.
The implementation of \thiswork is available at \url{https://github.com/AIS-SNU/smart-infinity}.
\end{abstract}

\section{Introduction}
\label{sec:intro}

Transformer~\cite{attention} based models currently dominate the natural language processing (NLP) field, effectively addressing the gradient vanishing problem that plagued recurrent neural networks (RNNs).
The transformer-based models tend to have large training parameters, still not showing overfitting problems to the training corpus~\cite{turing-nlg}. 
Therefore, the recent transformer-based model~\cite{gpt3, turing-nlg, t5, PaLM} has been linearly scaling the model size for the past few years, forming a group of large language models (LLMs).
As the models become larger, the major limiting factor becomes the GPU memory capacity. 
Often, dozens of GPUs are required, just to keep the training data on the memory, even when such computing power is not necessary such as fine-tuning~\cite{glue}.

One promising way to handle such a problem in a GPU memory-limited environment is storage-offloaded training~\cite{zero-infinity, stronghold}.   
For LLM training, it is known that the most memory-consuming factor is optimizer states and gradients~\cite{zero-offload, zero-infinity, stronghold, l2l, optimstore}, followed by model parameters and the activations.
Using this fact, the most widely adopted form of storage-offloaded training~\cite{zero-offload} is to store optimizer states and gradients in storage, while activations and model parameters are stored in host memory.
Because the memory capacities typically differ by orders of magnitude (e.g., $\sim$80G, $\sim$2TB, $\sim$100TB for GPU memory, host memory, and storage devices, respectively), they essentially serve as an extended memory hierarchy for GPUs.

Predictably, the use of large-capacity storage devices causes a severe bandwidth bottleneck because SSD storage devices only provide up to a few GB/s of bandwidth.
According to our study, more than 88\% of the total training time is consumed by transferring data from/to the storage.
At a glance, combining multiple storage devices using RAID0 solution could alleviate the problem.
However, such a method would fundamentally be bottlenecked by the limited number of PCIe lanes (or IO pins) of the host processor.
Moreover, PCIe lanes from CPUs nowadays are very valuable resources, which are shared by GPUs, FPGAs, NICs, or even system memories~\cite{cxl}.

In such circumstances, we aim to address these issues using computational storage device (CSD) products.
Based on decades of research~\cite{smssdwisconsin, smssducsc, activeflash, activedisk}, several commercial CSD products are available off-the-shelf~\cite{smssdsummit, ngd, scaleflux, eidetic}.
By placing computational engines near storage, they aim to offload computation of the host, and utilize the internal bandwidth of storage.
When more such devices are added, the available internal bandwidths linearly increase.

Utilizing these, we propose \thiswork, a fast LLM training system using CSDs.
To address the storage bottleneck problem, we mainly suggest moving the update task to the near-storage accelerator.
We identify that the optimizer states are only used in the update phase of the training but consume 75\% of the total storage bandwidth. 
By moving the update task to the custom accelerator inside CSDs, only the gradients and model parameters are required to be transferred, reducing 75\% of the entire traffic. 

There are several challenges to realizing \thiswork on an actual system. 
One issue is the CSD-internal data transfer, which causes detrimental effects on the system performance. 
To address this, we propose internal transfer handler optimization with buffer pre-allocation and swap overlapping techniques for drawing maximum throughput. 

Additionally, we propose a CSD-assisted gradient compression/decompression. 
When we scale the system with multiple CSDs, the bottleneck is at the host to storage traffic through the shared channel for sending gradients. 
To reduce this traffic, we compress the gradients on GPUs, and decompress them on the accelerator in CSD.
This greatly reduces the traffic toward better performance, without compromising the final model accuracy.

Most notably, \thiswork is integrated into PyTorch on a real system and is a ready-to-use framework.
Our evaluation shows that \thiswork achieves up to 2.11$\times$ speedup over the baseline.
The contributions are summarized as follows.
\begin{enumerate}
\item We propose \thiswork, a method to perform the update phase of LLM training in custom CSD accelerators. This greatly reduces the storage bandwidth bottleneck.

\item We propose an efficient data transfer handler structure to utilize storage bandwidth and hide the latency of CSDs.

\item We suggest a CSD-assisted gradient compression method to enhance the scalability of \thiswork. 

\item We integrate \thiswork on a real system with PyTorch to achieve up to 2.11$\times$ speedup on mixed-precision LLM training.
\end{enumerate}

\section{Background}
\label{sec:background}

\begin{figure}[t]
    \centering
        
    \includegraphics[width=\columnwidth]{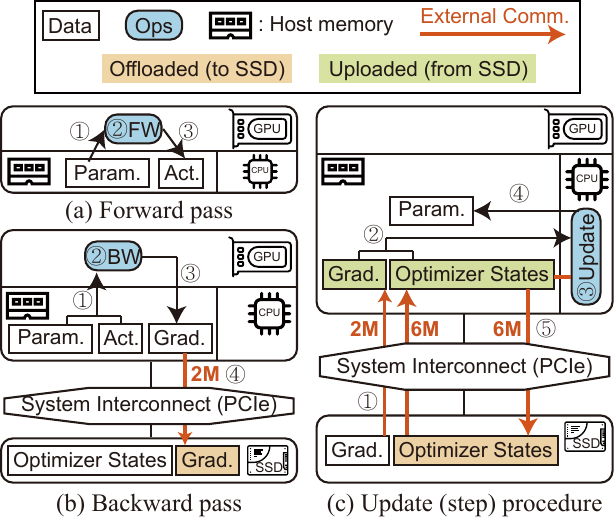}       
    \caption{
    A conceptual diagram of the storage-offloaded LLM training. Overview of 
 (a) the forward pass, (b) the backward pass, and (c) the update (step) procedure.
    }
    \label{fig:background}
\end{figure}

As discussed in \cref{sec:intro}, using aggregated GPU memory for entire LLM training is an expensive way.
A popular alternative is 
offloading solutions~\cite{zero-offload, zero-infinity, stronghold} utilizing the host memory or the storages in training LLMs with limited resources.  
We will describe them and discuss their limitations.

\subsection{ Overview of Dataflow in Storage-Offloaded LLM Training}

ZeRO~\cite{zero} analyzes the memory usage while LLM training and suggests methods to split the optimizer states (e.g., momentum and variance of Adam) for minimizing memory usage in distributed LLM training with multiple GPUs.
It points out the considerable additional memory requirements for optimizer states during LLM training. 

The FP32 optimizer states occupy 6$M$ capacity (model parameter, momentum, and variance in FP32) under mixed precision training~\cite{fp16}, regarding the FP16 model parameter size as $M$.
Considering that maintaining optimizer states in FP32 is an almost essential option for mixed precision LLM training~\cite{turing-nlg, ppt, efficientLLM}, this provides an insight that management of the optimizer state is a critical issue for LLM training in a resource-limited environment (e.g., lack of enough GPU memory).

Motivated by the above insight, many approaches~\cite{zero-offload, l2l, zero-infinity, stronghold} try to offload optimizer states of LLM training to host memory or storages, when GPU cannot hold the whole optimizer states due to memory capacity.
Host memory-offloaded training~\cite{zero-offload, l2l} provides the baseline concept of training LLMs using the host memory.

Because the computational intensity of updating optimizer states (element-wise operations) is low, it becomes attractive to offload the optimizer states to host memory and let the host CPU update parameters.
However, to train larger LLMs with model sizes of 345M~\cite{megatron-lm, gpt2} to 530B~\cite{gpt3, turing-nlg}, which cannot be trained even with the host memory-offloaded training, 
some works~\cite{zero-infinity, optimstore} provide solutions to further utilize storage devices.

On top of the host memory-offloaded training, storage-offloaded training~\cite{zero-infinity} suggests additionally using storage (usually NVMe) devices for training LLMs with limited hardware resources.
They offload the optimizer states to storages and make host memory hold only activations and mixed precision parameters.
It breaks the wall of the host memory size so that it can train even larger models on a single machine. 

\cref{fig:background} provides an overview of storage-offloaded training methods using mixed precision training. 
These methods split an LLM model into multiple blocks (e.g., layers), which have sizes that a GPU (or host memory) can handle at a time.
Before the training starts, the whole optimizer states for the model are initially stored in the storage (i.e., SSDs), as depicted at the bottom of \cref{fig:background}(b) and (c).
In the forward pass (\cref{fig:background}(a)), \circled{1} GPU loads the mixed precision parameters of a block.
\circled{2} After forward processing of the block in GPU, \circled{3} GPU checkpoints the activation of block to host memory.
For all blocks, the host iteratively conducts \circled{1}-\circled{3} and checkpoints all activations of the whole model.
In the backward pass (\cref{fig:background}(b)), for each block, \circled{1} GPU loads mixed precision parameters and activations, and \circled{2} conducts backward processing of the block in GPU, \circled{3} whose resulting gradients are sent to the host memory.
To minimize the usage of host memory space, \circled{4} the host offloads the created gradients to the storage device.

When the forward and backward passes are finished, the host starts the update procedure for the parameters of a model (\cref{fig:background}(c)) in a block-wise manner.
\circled{1} The gradients and optimizer states are uploaded from the storage to the host memory.
Then, \circled{2} The CPU loads the uploaded gradients and optimizer states, and \circled{3} updates them.
After the block update, \circled{4} the CPU replaces the mixed precision parameters with updated ones and \circled{5} offloads the optimizer states to the NVMe device.
The host keeps updating all blocks by repeatedly conducting \circled{1}-\circled{5}, and a forward pass of the next iteration follows after updating is finished.

From the help of storage devices, prior works enable the training of extreme-size LLMs when there is not enough GPU memory to hold the whole optimizer states.
However, as illustrated in \cref{fig:background}(b) and (c), storage-offloaded training suffers from a significant amount of upload/offload data transfers at every iteration, including the optimizer states (model parameter, momentum, and variance of $6M$ size in total) and
gradients handled in 32 bits by using a highly optimized offloading engine of \cite{zero-infinity} ($2M$).
Such data transfers pass through the system interconnect (PCIe) depicted with the red arrows in \cref{fig:background}.
\thiswork targets to reduce the data transfer through the system interconnect with a computational storage device (CSD).
\thiswork enjoys the aggregated internal CSD bandwidth instead of limited system interconnect bandwidth.
To retain such bandwidth with multiple CSDs, \thiswork utilizes the accelerators in CSDs while addressing some real system challenges towards efficient training.

\subsection{Computational Storage Devices}

\begin{figure}[t]
    \centering
        \includegraphics[width=.85\columnwidth]{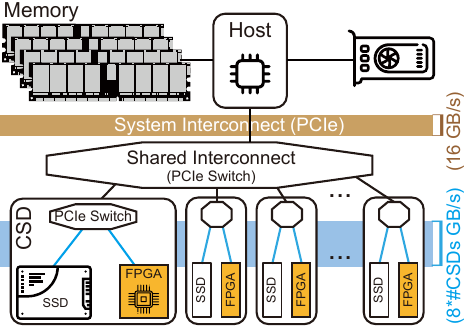}       
    
    \caption{
        An example environment with CSDs (e.g., SmartSSDs).
    }
    \label{fig:env}
\end{figure}

Computational storage devices, or near-storage processing (NSP) has been studied for years~\cite{smssdwisconsin, smssducsc, activeflash, activedisk}. 
By placing a computational engine closer to the storage devices, latency and bandwidth benefits can be obtained, in addition to reducing the host workload.
Among many CSD types~\cite{snia}, we specifically target ones shown in \cref{fig:env} where a lightweight FPGA accelerator is connected to SSDs via a switch inside the product, which is widely used for commercial products~\cite{smssdsummit, scaleflux, polardb, eidetic}.
\cref{fig:env} provides an example of a host environment with multiple such CSDs.

Generally, CSDs have two unique features compared to plain storage.
First, it has an accelerator (e.g., a lightweight FPGA) to compute data near storage.
Second, it has its own internal PCIe switch, providing a private inner path for direct peer-to-peer (P2P) communication between the SSD and accelerator without redundant storage-to-host and host-to-storage data traffic through the system interconnect.
A single CSD provides no bandwidth boost, as storage-to-FPGA and storage-to-host traffic transfers pass through the same number of PCIe lanes.  
However, 
when multiple CSD devices are on the system (\cref{fig:env}), the aggregated internal bandwidth linearly increases according to the number of CSDs, while the shared system interconnect bandwidth remains the same.
In \thiswork, we use SmartSSD~\cite{smssdsummit}, a representative commercially available CSD on the market.


%
%



\section{Motivation}
\label{sec:moti}

 \begin{figure}[t]
    \centering
    \includegraphics[width=\columnwidth]{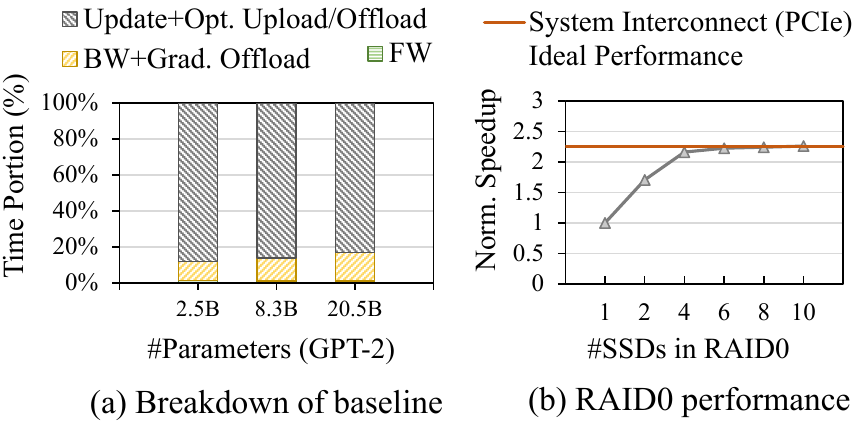}       
    \caption{
        (a) LLM storage-offloaded training time breakdown with various model sizes.
        (b) Speedup from the increasing numbers of SSDs using RAID0 solution.
    }
    \label{fig:motiv}
\end{figure}


We discussed the data transfer bottleneck between host memory and storage devices in storage-offloaded LLM training.
In this section, we directly analyze such an overhead with actual data and figure out why it is hard to be mitigated.

\textbf{Data transfer overhead.}
\cref{fig:motiv}(a) is the training time breakdown of LLM training with ZeRO-Infinity~\cite{zero-infinity}, state-of-the-art storage-offloaded training framework.
We conduct training in the environment with a single NVMe device (SSD).
For the detailed experimental environment, please refer to \cref{sec:env}.
As we described in \cref{sec:background}, we break down the storage offloaded training into 
forward (FW), backward (BW + Gradients Offload), and update (Update + Optimizer states Upload/Offload). 
Contrary to conventional training, the most time is spent on the update phase of over 80\% the training time, due to the storage access overhead.
The data transfer portion is significant regardless of model sizes, so it is a critical issue to be addressed in storage-offloaded LLM training.

\textbf{System interconnect bottleneck.}
One way to mitigate the data transfer overhead of NVMe devices is to make a higher communication bandwidth using RAIDs.
\cref{fig:motiv}(b) shows the normalized speedup of storage-offloaded training when increasing the number of SSDs using RAID0. 
Unfortunately, the speedup saturates after using more than four SSDs. 
Shared system interconnect becomes a new bottleneck when using more than four SSDs. 
Therefore, using the RAID solution for storage-offloaded LLM training has a limitation.

The motivational study demonstrates that data transfer overhead in storage-offloaded LLM training cannot be easily mitigated due to the limited resource of the existing system structure.
Therefore, when we use CSDs as an alternative, our primary goal is \textbf{minimizing the data transfer between storage devices and host memory through the shared system interconnect}, and we need to fully utilize {1) the aggregated internal bandwidth from multiple CSDs} and {2) the computational ability of the accelerator in each CSD}.

\section{\thiswork}
\begin{table}[t]
    \centering
    \caption{
        System interconnect traffic for storage-offloaded training with Adam optimizer.
    }
    \label{tab:traffic}
        
    \setlength{\tabcolsep}{3pt}
    \footnotesize
    \begin{tabular}{ccccc}
    \toprule

    Type & \multicolumn{2}{c}{Optimizer States} & \multicolumn{2}{c}{Gradients}\\
    \cmidrule(lr){2-3}
    \cmidrule(lr){4-5}
    SSD Operation & Read & Write & Read & Write \\
    \midrule
    ZeRO-Inf~\cite{zero-infinity}            & $6M$ & $6M$ & $2M$ & $2M$ \\
    \smartupdate        & $2M$ & $-$ & $-$ & $2M$ \\
    \SmartTopK (c\%)    & $2M$ & $-$ & $-$ & $c\% \times 2M$ \\
        \bottomrule
    \end{tabular}

\end{table}

\label{sec:smart_inf}
\cref{tab:traffic} provides an overview of changes in the system interconnect traffic from applying \thiswork.
First, to minimize the communication between storage devices and host memory, we offload the update computation from the CPU to the accelerator in CSDs (\SmartUpdate).
From this, we can benefit from the fast aggregate bandwidth of CSDs, reducing the communication overhead of existing storage-offloaded LLM training methods from $(6+2)M$ to $2M$.
Second, we propose a new internal data transfer handler structure with buffer pre-allocation and swap overlapping, which are directly applicable to \smartupdate.
The handler structure can be applied when integrating CSD to popular host codes (C++, Python), which is critical to the throughput of CSD applications in real systems.
Third, we propose a new method to further reduce remaining storage write traffic through the shared system interconnect from $2M$ to $c\% \times 2M$ ($c$: compression ratio) by compressing the gradients using the computational capability of FPGA in CSDs~(\SmartTopK).


\subsection{\SmartUpdate: Near-storage Update with CSD}
\label{sec:smartssd_update}

\begin{figure}[t]
    \centering
    
        \includegraphics[width=\columnwidth]{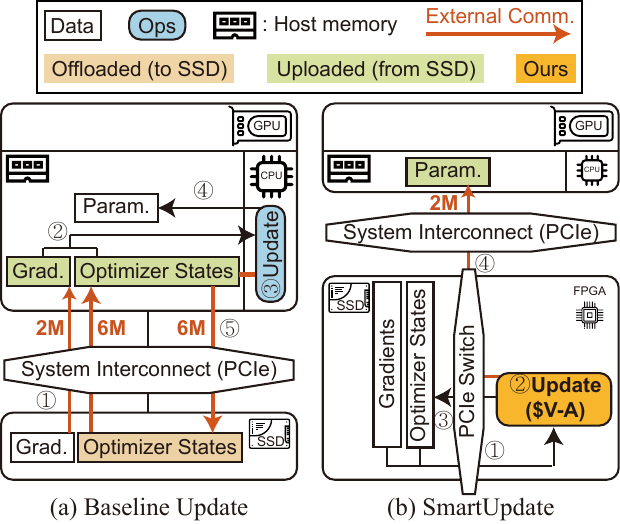}
    
    \caption{
        Update procedure of the storage-offloaded training with~(a)~baseline~\cite{zero-infinity} and (b) \smartupdate.
    }
    \label{fig:smart_offload}
\end{figure}

As discussed in \cref{sec:background}, the communication volume of the optimizer states in the update procedure is significant
in storage-offloaded LLM training. 
With conventional storage devices, this communication through the system interconnect (PCIe) is essential because the CPU conducts the update computation.
However, \smartupdate offloads the updating procedure from the CPU to the CSDs to reduce the interconnect communication volume.
The total data read from the storage is the same, but the aggregate bandwidth of multiple CSDs is utilized instead of the shared system interconnect.

\cref{fig:smart_offload} illustrates the detailed process of \smartupdate compared to the baseline~\cite{zero-infinity}.
We use Adam optimizer~\cite{adam} as an example because it is the primary choice for modern LLM training~\cite{bert, gpt2, gpt3}. 
As illustrated in \cref{fig:smart_offload}(a), the baseline updates parameters using the host CPU.
The gradients and optimizer states should be uploaded from SSDs to conduct the update procedure with the CPU.
After the update is finished, the optimizer states are offloaded to SSDs.
On the other hand, \smartupdate updates the parameters with the accelerator (i.e., FPGA) in a CSD.
Instead of uploading the gradients and optimizer states to host memory, \circled{1} \smartupdate directly loads them to FPGA through direct P2P communication which is possible because of the existence of an internal PCIe switch in each CSD.
After the loaded gradients and optimizer states are stored in the accelerator memory, \circled{2} the accelerators update the parameters using an update module (orange box).
We discuss the detailed architecture in a separate section (\cref{sec:updater}).
\circled{3} While the baseline offloads the optimizer states from the host memory to SSDs after updating, \smartupdate directly sends back the updated optimizer states to SSD through internal P2P communication.
\circled{4} The updated weight parameters ($2M$) are transferred from SSD to the host memory.
This traffic did not exist in the baseline, but it is comparably small overhead considering the size of the previous traffic volume was upload/offload optimizer states ($6M$, three single-precision variables per parameter). 
In addition, this upstream traffic can be overlapped with update steps for other parameters.

When quantifying the total communication volume, the baseline requires $8M$ (gradients + three optimizer states) for both read and write. 
\smartupdate minimizes the total communication volume through the system interconnect to only $2M$ read during the update phase for the updated parameters and $2M$ write during backward for the gradients.

When \smartupdate is used with a single CSD, the bottleneck simply moves from the system interconnect to the CSD-internal switch.
However, the real advantage appears when scaling the number of CSDs.
When the number of CSDs increases, the aggregate bandwidth between the FPGA and SSD increases linearly, while that of the system interconnect (PCIe) to the host stays constant.
While there is a slight benefit of increased computational capability from the increased number of FPGAs, the aggregated internal bandwidth is the main driver for the speedup of \thiswork. 

\subsection{Internal Data Transfer Handler for \smartupdate}
\label{sec:opt}

\begin{figure}[t]
    \centering
    \includegraphics[width=\columnwidth]{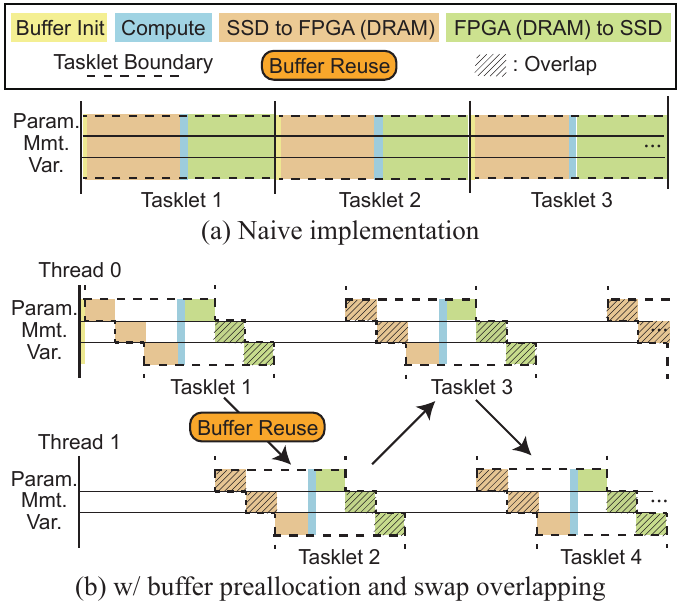}       
    \caption{
        (a) Naive implementation of \smartupdate. (b)~\smartupdate with internal data transfer handler optimization. 
    }
    \label{fig:opt_tech}
\end{figure}

 As described in \cref{sec:background}, a single iteration of storage-offloaded training splits the model into multiple blocks and processes a single block at a time. 
The number of parameters in each block is decided based on the estimated memory requirement.
Similarly, \smartupdate also sets the size of a \emph{subgroup} of model parameters according to the accelerator's device memory capacity, and each subgroup is processed with a single \emph{tasklet}.
In a naive implementation shown in \cref{fig:opt_tech}(a), the tasklets are executed sequentially because each tasklet will occupy the entire memory space.
Fortunately, we found that there is room for throughput enhancement by overlapping data transfer between subgroups. 
However, naive overlapping of data transfers between subgroups requires additional device memory space to direct P2P communication, which leads to out-of-memory (OOM) errors in device memory. 
Therefore, we propose an internal data transfer handler optimization which leads to better throughput of \smartupdate, addressing the device memory consumption issue.

The key idea of the proposed handler is to separate the buffer for variables and lazily transfer non-urgent variables. 
In a naive implementation (\cref{fig:opt_tech}(a)), the procedure is as follows: the buffers for current subgroup are allocated, the variables are read from the SSD, parameters are updated (light blue bars), and the variables are written back to the SSD.
Afterward, the buffer is deallocated, so a buffer for the next subgroup can be allocated.
 With our optimization technique (\cref{fig:opt_tech}(b)), the device memory buffer is preallocated at initialization of \smartupdate, separately for the largest possible size of each optimizer state variable.
We dedicate two threads (thread 0 and 1) running at the CPU to manage the allocated buffers.
Initially, thread 0 owns the buffers. 
When it finishes the update computation, it immediately writes the model parameters back to the SSD, because it is needed by GPUs to conduct forward/backward passes.
However, the other variables (e.g., momentum and variance) are not urgent to be written back, because they will be used only at the update phase of the next training iteration.
Thus, thread 0 defers the writeback of the remaining variables and signals thread 1 to start loading the model parameters of the next subgroup. 
Because we have preallocated the buffer for the largest subgroup size, thread 1 can directly reuse the same buffer. 
For the remaining variables, the writeback of thread 0 and loading of thread 1 are performed similarly by reusing the buffers, but in a lower priority because it is not critical for the forward/backward phases.
Through the optimization, we obtain several benefits:
1) avoid reallocation of device memory, 2) the GPU can start forward/backward phases earlier, and 3) the data transfers to the SSDs are overlapped.

\subsection{\SmartTopK: CSD-aided Gradient Compression}
\label{sec:smartssd_topk}


\smartupdate reduces a significant amount of the SSD traffic by removing the transfer of optimizer states to the host.
Given multiple CSDs in the system, the optimizer states ($6M$) are transferred within the CSDs through the internal bandwidth linearly scaled by the number of CSDs. However, gradients ($2M$) still go through the system interconnect, which becomes a new bottleneck when the number of CSDs increases.
Unfortunately, it is difficult to overlap gradient offloading with the update step because there are some constraints for mixed precision LLM training before starting the update step.
First, not-a-number (NaN) and infinity value (Inf) due to the limited range of half-precision must be checked before the update for loss scaling~\cite{fp16}.
Second, the norm of total gradients from the whole model is required for gradient clipping before the update phase.
Moreover, the write bandwidth is often far lower than that of the read of SSDs, which aggravates the issue.

Fortunately, the fact that bottleneck is caused by gradients provides an interesting opportunity to \thiswork: compressing gradients. 
It is widely known that DNN training is tolerable to some degree of errors, especially the gradients. 
Gradient compression methods~\cite{adacomp, dgc, powersgd, scalecom, optimus_cc} are widely used to mitigate communication overhead in neural network training, and it is almost a norm to train modern larger models in distributed settings~\cite{dalle, optimus_cc}.

\begin{figure}[t]
    \centering
     
    \includegraphics[width=\columnwidth]{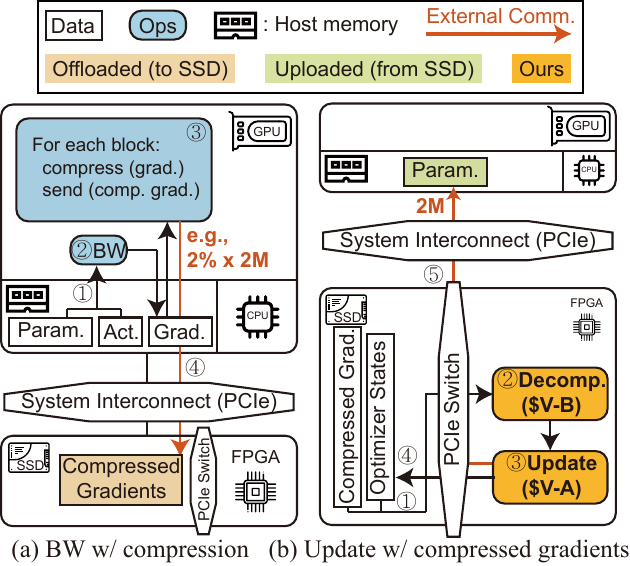}       
    \caption{
        An overview of \SmartTopK. (a) The remaining gradient offloading overhead in \smartupdate is further reduced by gradient compression. (b) FPGA in CSD conducts the decompression of the compressed gradients.
    }
    \label{fig:smart_topk}
\end{figure}

With the aid of the accelerator in CSDs, we propose \SmartTopK that applies gradient compression on top of \smartupdate: compress the gradients using GPU, and decompress them before updating on CSDs. 
While there are many variants of algorithms, we implement the magnitude-based compression method~\cite{adacomp, dgc} for \SmartTopK.
In the method, the gradients are first sorted by their magnitude, and higher ones are chosen.
The lower-magnitude gradients are replaced by zero because they have a relatively small impact on the original gradient direction compared to the higher-magnitude ones.
The compression results are indices list and values list, each representing positions of the high magnitude gradients and corresponding values.
%
This is a viable choice because the relatively heavier job or sorting can be done in powerful GPUs, and the lightweight FPGAs in CSD perform the decompression part that only involves scattering the values according to the index.
Another approach, such as low-rank decomposition~\cite{powersgd, gradzip}, could be applied to \SmartTopK.
However, tuning the floating-point matrix multiplication performance is challenging and non-trivial~\cite{autosa, polysa, hls }.
The magnitude-based compression provides comparable model quality 
at low DSP cost, so we settled on the magnitude-based compression.

\cref{fig:smart_topk} illustrates the overall process of \SmartTopK.
After the backward pass generates the gradients (\circled{1} and \circled{2} of \cref{fig:smart_topk}(a)), \circled{3} GPU compresses the gradients.
For example, GPU selects the higher magnitude (e.g., 1\%) elements by sorting each block.
Instead of offloading the full gradients vector to SSD, \circled{4} only the compressed gradients are offloaded to SSDs.
In the update procedure (\cref{fig:smart_topk}(b)), through the internal CSD P2P communication, \circled{1} the compressed gradients and optimizer states are loaded to FPGA.
Because the full matrices are required for the element-wise aspect of the update operation (e.g., Adam~\cite{adam}), we first \circled{2} decompress the compressed gradients to the same size as the original gradient vector using index information before the update.
For this process, we designed the decompression module (orange box), which will be described in a separate section (\cref{sec:decompressor}).
After the decompression, \circled{3}-\circled{5} \SmartTopK follows the same procedure as illustrated in \cref{sec:smartssd_update}.
With \smartupdate and \SmartTopK, the remaining data transfers through the system interconnect now become the offloading compressed volume of gradients and upstream updated weight parameters, which are essential information to proceed to the next forward and backward pass in GPU.
Combining the facts that the weight parameters are more urgently passed with our internal data transfer handler~(\cref{sec:opt}) and SSDs are faster on reads, \SmartTopK greatly helps to scale to the larger number of CSDs on \thiswork.





\subsection{Workload Distribution to Multiple CSDs}


In the previous sections, the details of \thiswork were provided using a single CSD for easy understanding.
However, \thiswork achieves acceleration from using multiple CSDs, utilizing the high internal bandwidth. Therefore, it is meaningful to discuss how \thiswork manages them.

The key idea of allocating the workload of each CSD is that all the operations in updating optimizer states are conducted in an element-wise manner.
Using this, \thiswork flattens the model parameters and equally distributes them to the CSDs,
where each CSD takes the responsibility to update the owned parameters.
Then, the optimizer states are allocated and initialized at CSDs that own the associated parameters.
After a backward pass, the generated gradients are offloaded to the owner CSD, who conducts update computation via P2P communication between its attached FPGA and SSD. 

Because of using flattened model parameters, the distribution procedure is agnostic to the model architecture. 
This allows for a simple adoption of \thiswork, where end users do not need to consider the model architecture information such as the layers, hidden dimensions, or number of heads.



\section{Accelerator Architecure}
\label{sec:arch}

\begin{figure}[t]
    \centering
    \includegraphics[width=\columnwidth]{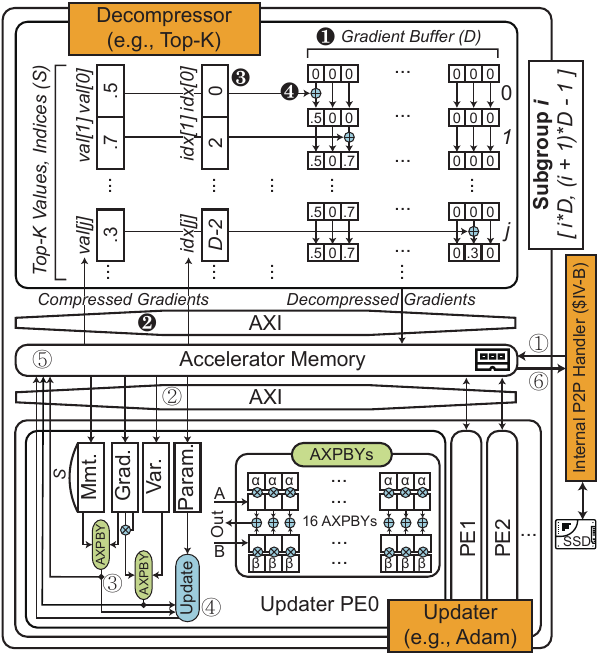}   
    \caption{
        \thiswork microarchitecture.
    } 
    \label{fig:arch}
\end{figure}

In this section, we introduce the accelerator architecture implemented in the FPGA to support \smartupdate and \SmartTopK. 
\cref{fig:arch} shows the microarchitecture of the updater for \smartupdate (\cref{sec:smartssd_update}) and decompressor for \SmartTopK (\cref{sec:smartssd_topk}).
Even though \thiswork mainly describes using Adam optimizer and magnitude-based compression method, 
the accelerator is designed to support various algorithms of user's choice.


Overall, \thiswork's updater and decompressor processes the model in units of a subgroup that fits into the DRAM size of the accelerator ($D$).
Therefore, when processing subgroup $i$, the parameters with the indices from  $i\times D$ to $(i+1) \times D-1$ are the target parameters. 
$S$ indicates the processing chunk size that fits into an internal BRAM buffer of the accelerator.

\subsection{General Updater}
\label{sec:updater}
The lower part of \cref{fig:arch} shows the example updater for Adam optimizer.
For the updater, \thiswork \circled{1} loads the subgroup size ($D$) amount of the gradients, optimizer states, and target parameters using the optimization techniques illustrated in \cref{sec:opt}.
The updater consists of multiple updater processing elements (PEs), which update the target parameters in parallel.
\circled{2} Each PE gets the gradients, the optimizer states, and the target parameters from the accelerator memory.
Generally, optimizer algorithms require various types of moving averages.
Therefore, we placed SIMD \textit{AXPBY} units, which can handle general averaging operations.
It takes two input vectors (i.e., $A$ and $B$) and multiplies the dedicated coefficients (i.e., $\alpha$ and $\beta$) to them.
After multiplication, it conducts the element-wise addition of two vectors.
The above procedure can be used for various averaging operations by changing the coefficients.
\circled{3} Using the AXPBY units, the updater incorporates the optimizer states (e.g., momentum and variance) into the gradients.
\circled{4} After all the optimizer states are applied to gradients, the updater updates the target parameter using them.
\circled{5} The new optimizer states, and the updated parameters are passed to the accelerator memory. 
\circled{6} The updated data are swapped out to the storage of CSD via the data transfer handler of \cref{sec:opt}.
Note that the updater can be extended to other optimizers such as SGD, AdaGrad~\cite{adagrad}, and AdamW~\cite{adamw} because most optimizers use variations of moving averaging.
We further implemented and tested other updaters in \cref{sec:other_optimizers}.

\subsection{General Decompressor}
\label{sec:decompressor}
As a representative compression method, \thiswork applies a magnitude-based (i.e., Top-K) algorithm.
The upper part of \cref{fig:arch} is the microarchitecture of the example decompressor for Top-K compression.
\bcircled{1} At the first iteration, the \textit{gradient buffer} for the target parameters is initialized with zero.
\bcircled{2} The decompressor gets the compressed gradients from the accelerator memory.
A typical Top-K compression algorithm compresses the gradients into two parts, indices and values.
Therefore, the decompressor loads the buffer size ($S$) amount of indices and values iteratively.
For each iteration, \bcircled{3} the decompressor figures out the target position using the value of indices ($idx[j]$) to map the values ($val[j]$).
\bcircled{4} In the output vector, the partial Top-K elements of the target parameters selected in this iteration $j$ are filled with their values, and the others are filled with zero.
Until decompression finishes, the decompressor repeats \bcircled{2}-\bcircled{4}.
When all Top-K elements are processed, the gradient buffer contains the vector filled with Top-K gradient values dedicated to the target parameters of subgroup $i$.
After decompression completes, decompressed gradient values are now ready in the accelerator memory, so that the updater (\cref{sec:updater}) can use them.

\section{Implementation}
\label{sec:impl}
\begin{figure}[t]
    \centering
    \includegraphics[width=\columnwidth]{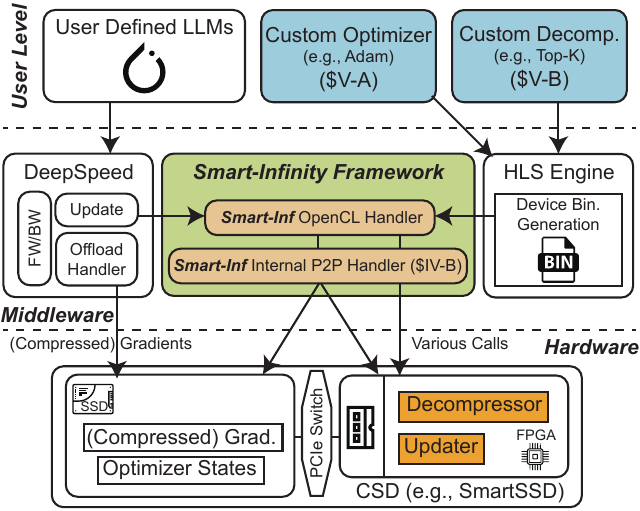}       
    \caption{An overview of \thiswork, which is a flexible ready-to-use framework.
    Users can apply their own optimizers and decompressor HLS modules with minimal modifications.
    }
    \label{fig:impl}
\end{figure}

This section describes the main software components of \thiswork on top of DeepSpeed~\cite{DeepSpeed_github}, an open-source framework that is highly optimized and widely used for LLM training with PyTorch. 
We mainly focus on replacing the parameter update feature of DeepSpeed ZeRO-Infinity~\cite{zero-infinity}. 
To build \thiswork into a callable module, we exploit \textit{disutils}~\cite{disutils} to build C/C++ implementations as additional modules of Python.
Because of this, \thiswork can be enabled by simply specifying an option, and automatic compilation starts when the DeepSpeed engine is initialized. 
This indicates that any LLM training codes implemented with DeepSpeed can be run with \thiswork with no modification, making \thiswork a practical and powerful framework.
Note that wrapping any native PyTorch LLM training code with DeepSpeed is also straightforward and requires little effort.

\cref{fig:impl} \emph{User Level} shows the design flow for users to customize \thiswork decompressor/updater. 
Users can freely modify the customized logic for \thiswork via high-level synthesis (HLS) codes for the decompressor/updater. 
\thiswork provides HLS templates for implementing their own compression or weight update logic. 
The templates also consist of a performance analyzer and a sanity checker of logic.

\cref{fig:impl} \emph{Middleware} describes interaction between \thiswork and DeepSpeed runtime engine. 
DeepSpeed supports forward/backward execution in mixed precision. The generated gradients are offloaded during backward execution. 
We carefully modify the gradient offloading path because the gradients should be located in the corresponding SSD to execute the update in FPGA via internal P2P communication.
After updating parameters with \thiswork, the updated parameters should be passed to the DeepSpeed runtime engine.
\thiswork engine can directly communicate \textit{`torch.Tensor'} type with PyTorch application using pybind11~\cite{pybind11}, a lightweight library that exposes Python types into C++.

\cref{fig:impl} \emph{Hardware} illustrates how \thiswork engine interacts with the hardware components.
\thiswork uses the Xilinx OpenCL extension~\cite{xilinx_opencl} to interact with the attached FPGA in runtime. 
When initializing the device with OpenCL, it is critical to figure out which FPGA device is directly connected via its internal PCIe switch to the specific SSD.
As described in \cref{sec:opt}, the internal data transfer handler calculates the required device buffer size and initially pre-allocates OpenCL buffer using \textit{`CL\_MEM\_EXT\_PTR\_XILINX'} flag at once.
Then, the buffer can be re-used for direct internal P2P communication with the attached NVMe SSD.
Some standard file access Linux system calls are supported for direct P2P data transfer between SSD and FPGA device memory.
We use \textit{pread/pwrite} system call to the P2P buffer, which operates the direct internal P2P communication feature of SmartSSD.

\section{Evaluation}
\label{sec:exp}

\subsection{Experimental Environment}
\label{sec:env}

\begin{table}
\footnotesize
\centering
\caption{Experimental environment.}\label{tab:environment}
 \def\arraystretch{1.0}
         
\resizebox{\columnwidth}{!}
{
\begin{tabular}{ccc}
\toprule
\multirowcell{7}[-0.4ex]{\textbf {HW}} 
&GPU & NVIDIA A5000, \rev{A100 (40GB), A4000 }  \\
\cmidrule(lr){2-3} 
&CPU & Xeon(R) Gold 6342, 2$\times$48C 96T  \\
\cmidrule(lr){2-3} 
&Memory& 32$\times$32GB DDR4-3200\\
\cmidrule(lr){2-3} 
&SSD & SAMSUNG SmartSSD, 4TB   \\
\cmidrule(lr){2-3} 
&PCIe Expansion & H3 Falcon 4109  \\

\midrule

\multirowcell{8}[-0.8ex]{\textbf {SW}} 
& OS & Ubuntu 20.04 LTS \\ 
\cmidrule(lr){2-3} 
& Python / PyTorch   & 3.9 / 1.12.1\\ 
\cmidrule(lr){2-3} 
& CUDA / OpenCL & 11.6.2 / 2.2 \\ 
\cmidrule(lr){2-3}
&Vitis / XRT & 2023.1 / 2.12.427 \\ 
\cmidrule(lr){2-3} 
& Model & GPT-2, BERT, BLOOM, ViT \\ 
\cmidrule(lr){2-3} 
& Deepspeed & 0.9.3 \\ 

 \bottomrule
\end{tabular}
}

\end{table} 

\begin{figure*}[t]
    \centering
    \includegraphics[width=.95\textwidth]{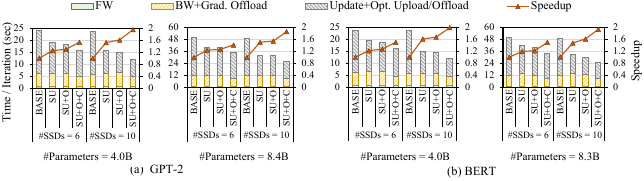}       
    \caption{
        Training time breakdown and speedup of \thiswork over the baseline (BASE).
        SU indicates \smartupdate, and SU+O means the optimized \smartupdate with internal data transfer handler.
        SU+O+C uses \SmartTopK on top of SU+O.
    }
    \label{fig:main_speedup}
\end{figure*}

The overall experimental environments are shown in \cref{tab:environment}.
To evaluate the efficiency of utilizing CSD, we use at most 10 SAMSUNG SmartSSD~\cite{smssdsummit} devices.
Each SmartSSD has a 4TB NVMe SSD which directly communicates with Kintex UltraScale+ KU15P FPGA through a PCIe Gen3.0~x4 bus.
The attached FPGA has approximately 522K LUTs, 984 BRAMs, 1968 DSPs, and DDR4 4GB DRAM.
We use the NVMe SSD of SmartSSD for a fair comparison with the baseline.
We compose software RAID via Linux \textit{mdadm}.
We connect SmartSSDs via a PCIe expansion~\cite{falcon}, which is helpful to increase the physical slots of the whole system when the PCIe lanes are limited.
We further discuss the storage expansion in \cref{sec:expansion}.
For LLM training, we equipped RTX A5000 (24GB), Tesla A100 (40GB), and RTX A4000 (16GB), which are widely used GPUs to train DNNs. We use RTX A5000 as a default if it is not stated otherwise.

As described in \cref{sec:background}, DeepSpeed ZeRO-Infinity~\cite{zero-infinity} is highly optimized for overlapping data transfer in storage-offloaded training and widely used in mixed-precision training.
Additionally, it has internal implementation with AVX operations to provide faster updates with CPUs.
Therefore, we set DeepSpeed ZeRO-Infinity~\cite{zero-infinity} with software RAID0 as the baseline (\textbf{BASE}).
In the following sections, we notate \smartupdate as \textbf{SU} (\S\cref{sec:smartssd_update}), \smartupdate with the optimization techniques of internal data transfer handler as \textbf{SU+O} (optimized SU, \S\cref{sec:opt})), and optimized \smartupdate with \SmartTopK as \textbf{SU+O+C} (\S\cref{sec:smartssd_topk}).
If not stated, the default compression ratio is 2\% (1\% gradients with indices), which indicates that it communicates only 2\% of the original communication volume. 
And we used batch size 4 as our default setting, since we targeted the situation where GPU memory size is limited.
In addition, we mainly chose two different language models (GPT2~\cite{gpt2}, BERT~\cite{bert}), which are representative decoder/encoder-only models to measure speedup and accuracy. Further, we conduct experiments in \cref{fig:model_sensi} for the other two models (BLOOM~\cite{bloom}, ViT~\cite{ViT})

\subsection{Implementation Results} 
\begin{scriptsize}
\begin{table}
    \caption{
        Resource utilization of implementation results for Adam updater and Adam updater with Top-K decompressor.
    }
    \label{tab:imple_result}
    \resizebox{\columnwidth}{!}
    {
    \setlength{\tabcolsep}{3pt}
    \begin{tabular}{cccccccc}
    \toprule
         & Module &  \multicolumn{1}{c}{ LUT (522K)}  & \multicolumn{1}{c}{BRAM (984)} & \multicolumn{1}{c}{ URAM (128)} & \multicolumn{1}{c}{DSP (1968)} \\
         
         \midrule
          & Adam  & 33.66\%  & 27.13\%  & 34.38\%  & 11.03\% \\
          & Adam w/ Top-K & 34.12\%  & 27.13\%  & 35.94\%  & 11.03\% \\
        \bottomrule
    \end{tabular}
    }
\end{table}
\end{scriptsize}

\cref{tab:imple_result} shows the resource utilization of the implementation results of the microarchitectures on the FPGA, for the Adam optimizer and Top-K decompressor.
Note that in the actual usage of \thiswork framework, the user does not always need to implement the modules manually except when they need customized logic.
\thiswork provides general templates for generating the device binary, as discussed in \cref{sec:impl}.
For implementing Adam, the updater consumes roughly a quarter of the available resources for implementing floating point AXPBY vector arithmetic and pipeline registers.
On the other hand, the decompressor consumes a small amount of logic, because the Top-K decompressor only requires routing the value to the right location without any arithmetic. 
There is much room left for extra logic despite the FPGA being lightweight, possibly allowing other extensions of applications.

\subsection{Performance Comparison and Ablation}


\cref{fig:main_speedup} shows the training time breakdown and the speedup of \thiswork compared to the baseline with ablation.
We break down the training time into three parts, forward time, backward with gradient offload time, and update with the optimizer states upload and offload time.
Overall, the baseline training time results suffer from the severe communication overhead as discussed in \cref{sec:background} and \cref{sec:moti}.
In GPT-2 8.4B model with 6 SSDs, the update time with the optimizer states communication time consumes 75.57\% of the training time of the baseline.
Even when the number of SSDs increases to 10, the training time stays constant due to the system interconnect bottleneck, and this trend persists for all test cases.
\smartupdate (SU) reduces this communication volume through the system interconnect, providing 1.18$\times$$\sim$1.24$\times$ speedup with 6 SSDs, and  1.54$\times$$\sim$1.60$\times$ speedup with 10 SSDs.
Further applying the transfer handler optimization (SU+O), 
the speedup from \smartupdate is boosted to up to 1.60$\times$$\sim$1.66$\times$ with 10 SSDs.
This result suggests that the optimization techniques successfully overlap the SSD-FPGA communications.
Additionally, on the top of the optimized \smartupdate, \SmartTopK provides 1.22$\times$$\sim$1.31$\times$ additional speedup over \smartupdate, showing 1.85$\times$$\sim$1.98$\times$ speedup over the baseline.
Through the breakdown results, we can find out that \SmartTopK successfully reduces the remaining gradient offloading time of the optimized \smartupdate further.
For all models tested, the speedup trend is almost identical.
This is because the modern LLM models are all based on Transformers~\cite{attention} and only differ in some model design parameters.
Furthermore, in all cases, the main bottleneck is the storage bandwidth, making the model size or structure relatively less important for the training time. 
This explains the constant speedup trends.

\subsection{Scalability to Larger Models}

\begin{figure}[t]
    \centering
    \includegraphics[width=\columnwidth]{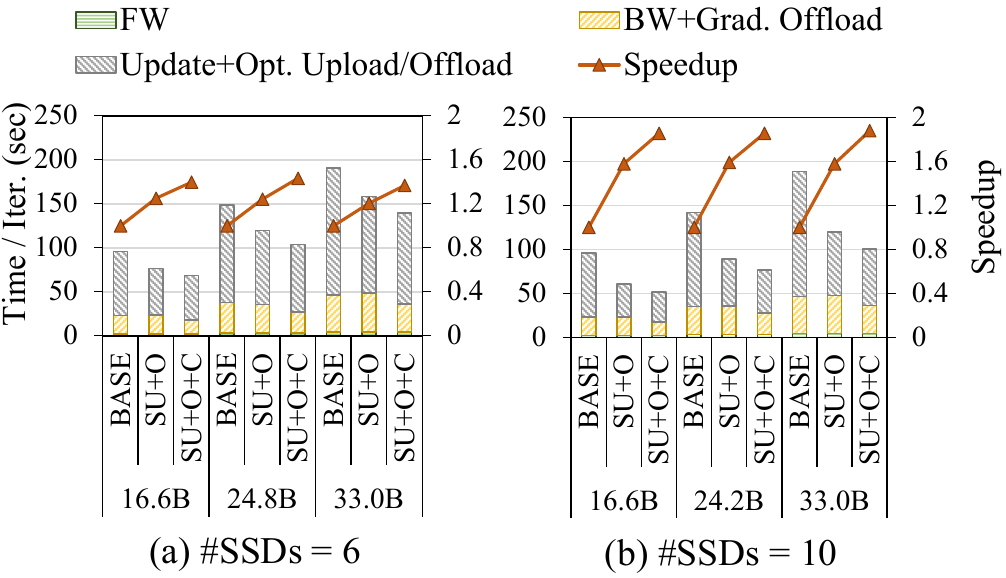}       
    \caption{
        Scalability on larger model sizes (16.6B to 33.0B) of \thiswork compared to the baseline.
    }
    \label{fig:model_scale}
\end{figure}

Storage-offloaded training allows training much larger models using limited resources.
Therefore, it is essential to check whether \thiswork brings stable speedup over the baseline on the various large model sizes.
\cref{fig:model_scale} is the scalability test of \thiswork on larger GPT models with various sizes compared to the baseline.
The results show that \thiswork achieves consistent speedup with larger models. 
Even with GPT-2 33.0B model, \thiswork provides 1.37$\times$ and 1.88$\times$ speedup over the baseline using 6 SSDs and 10 SSDs, respectively.
The speedups are stable on large models because the communication portion of the training time is maintained.
It is natural because the communication volume is proportional to the number of model parameters in transformer model training.
Additionally, using more number of CSDs still brings speedup to the larger models.
Therefore, \thiswork is scalable on larger models while the baseline suffers from the huge uploading and offloading overheads.

\subsection{Experiments on the Number of CSDs and GPU Grade}

\begin{figure}[t]
    \centering
    
        \includegraphics[width=\columnwidth]{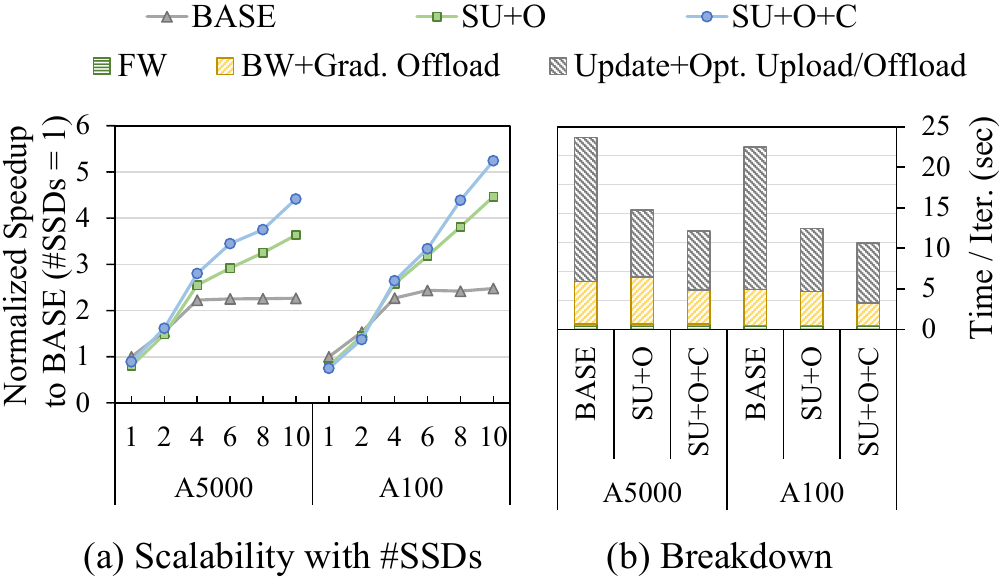}       
    
    \caption{ 
       (a) Scalability with \#SSDs of \thiswork compared to the baseline in  A5000 and A100 GPU settings. (b) Training time breakdown of \thiswork and the baseline in both GPU settings with ten SSDs.
    } 
    
    \label{fig:csd_scale}
\end{figure}

\thiswork benefits from the aggregated bandwidth of CSDs, so it is essential to check the sensitivity of speedup on the number of CSDs.
Additionally, it is helpful to check the speedup when using a higher-end GPU.
In \cref{fig:csd_scale}(a), we scaled the number of CSD from one to 10 while keeping the model as GPT-2 4.0B with the default A5000 GPU and the higher-end A100 GPU.
In both GPU setups, the baseline does not scale beyond 4 SSDs, whose aggregate SSD bandwidth reaches roughly that of the PCIe system interconnect bandwidth.
On the other hand, \thiswork mainly depends its speedup on the aggregate internal bandwidth, not the system interconnect. 
Because of this, \thiswork shows almost linear speedups along with increasing the number of CSDs.
On a single CSD, there is a slight slowdown, which is expected. 
It is because there is no bandwidth increase with a single CSD, and the base system overhead to utilize the FPGA adds a slight overhead.
When utilizing a higher-end GPU (i.e., A100), the portion of computation (FW and BW) becomes smaller.
Therefore, the portion of data transfer gets larger, and this causes the speedups in the A100 GPU setting to be generally higher than the ones in the default A5000 setting.
Additionally, \cref{fig:csd_scale}(b) breaks down the training time when using ten SSDs in both settings, and it shows that \thiswork still provides up to 2.11$\times$ speedup in the A100 GPU setting.

\subsection{Extension to Other Optimizers}
\label{sec:other_optimizers}

\begin{figure}[t]
    \centering
    \includegraphics[width=\columnwidth]{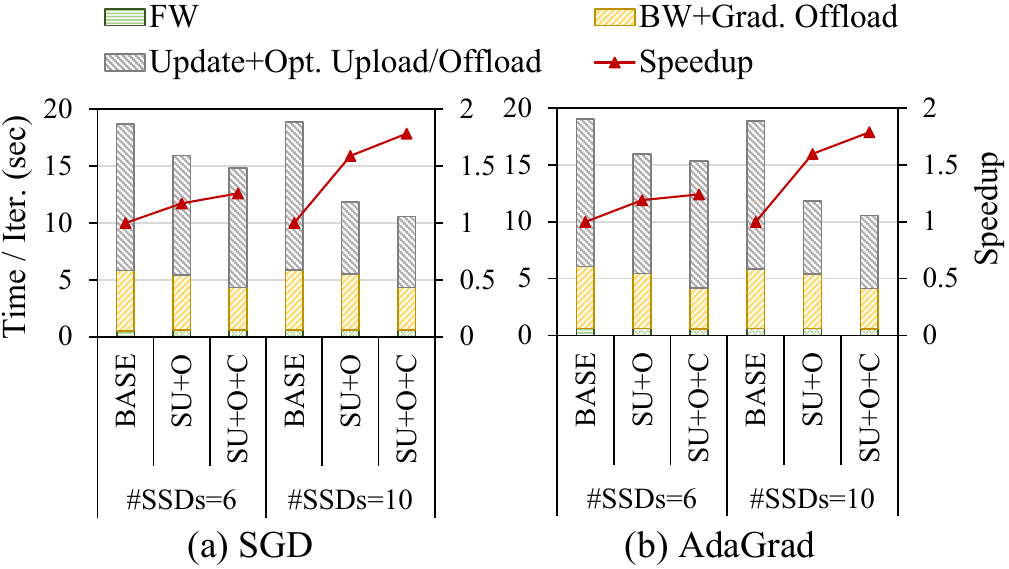}      
    
    \caption{
        Applying \smartupdate to other optimizers.
    }
    \label{fig:other_optimizers}
\end{figure}

\thiswork provides a general updater that can support various optimizers.
Besides the default Adam updater, we additionally implemented the updaters for stochastic gradient descent (SGD) with momentum and AdaGrad~\cite{adagrad}.
\cref{fig:other_optimizers} shows the speedup of \thiswork over the baseline when changing the optimizer type in GPT-2 4.0B.
SGD requires $3/4 \times$ less optimizer states because it uses three states (parameter, gradient, and momentum) instead of four.
Therefore, the speedup when using SGD becomes slightly lower than that of Adam.
AdaGrad also incurs $3/4 \times$ offloading volume (parameter, gradient, and variance) compared to Adam, so the speedup is similar to the SGD.
Recent optimizers generally use a similar or larger number of optimizer states, indicating \thiswork would be an efficient way for other optimizers. 

\subsection{Applying to Other Models}

\begin{figure}[t]
    \includegraphics[width=\columnwidth]{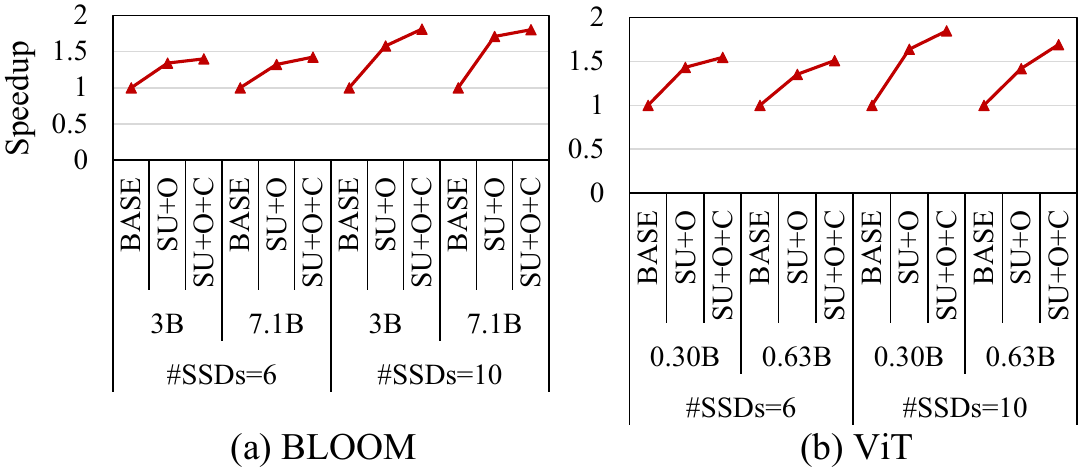}
    \caption{
        Application of \thiswork to BLOOM and Vision Transformer (ViT).
    } 
    \label{fig:model_sensi}
\end{figure}

\thiswork can be generally applied to other transformer-based models with heavy memory overhead.
To check such applicability, we tested the speedup of \thiswork on BLOOM~\cite{bloom} and vision transformer (ViT~\cite{ViT}) in \cref{fig:model_sensi}.
BLOOM is another open-sourced multilingual LLM.
ViT is a representative transformer-based model for vision-related tasks. They are chosen to demonstrate that \thiswork is not limited to specific model sensitive aspects.
On such models, \thiswork stably shows 1.32$\times$$\sim$1.85$\times$ speedup, similar to the results we have shown for GPT-2 and BERT.

\subsection{Accelerator Throughput Analysis}

\begin{figure}[t]
    \centering
    \includegraphics[width=\columnwidth]{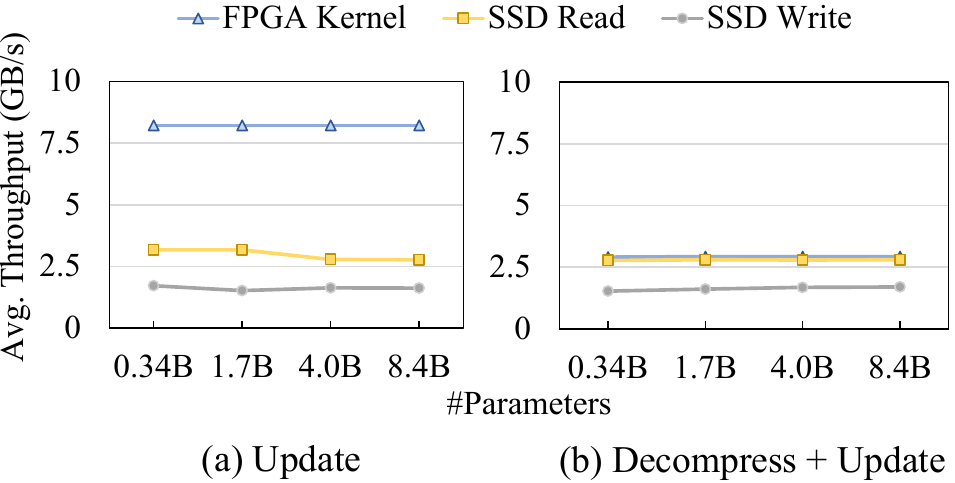}       
    \caption{
        Computational throughput of \thiswork's modules compared to NVMe SSD read and write performance.
    }
    \label{fig:throughput}
\end{figure}

For achieving high performance with \thiswork, it is crucial that the throughput of the updater and the decompressor modules keep up with the SSD bandwidth.
Therefore, we compared the throughput of updater, decompressor, and SSD read/write in \cref{fig:throughput}.
The throughput of the updater is above 7GB/s, which is sufficiently higher than the SSD read/write.
The decompressor also slightly surpasses the throughput of the SSD read. 
In principle, it could be possible to obtain higher decompression speed by deploying more such engines considering the FPGA resource utilization.
However, we chose to save the FPGA resources for later extensions (\cref{sec:dis_comp}) as it was not slowing down the system.

\subsection{System Cost Efficiency Analysis}

\begin{figure}[t]
    \centering
        \includegraphics[width=\columnwidth]{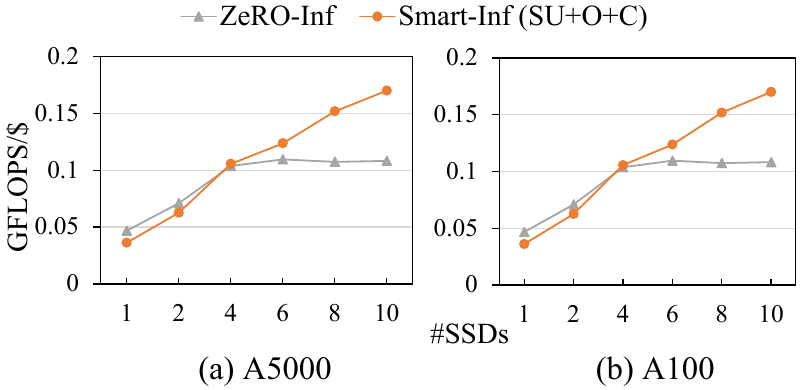}      
    \caption{ 
       Performance/$\$$ of the baseline and \thiswork.
    } 
    \label{fig:system_cost}
\end{figure}

It is meaningful to directly compare the system cost efficiency of \thiswork with the baseline storage-offloaded training~\cite{zero-infinity}.
In \cref{fig:system_cost}, we analyze their system cost compared to GFLOPS/$\$$ in the setting used for \cref{fig:main_speedup} (GPT-2 4.0B).
The system costs include the server cost (around $\$$45,000, including CPU, RAM, PCIe expansion, etc.), storage cost, CSD (i.e., SmartSSD) cost, and GPU cost (around $\$$2,000, $\$$7,000 for A5000 and A100, respectively).
The cost of a SmartSSD is around $\$$2,400, which is 6$\times$ more expensive than the same capacity SSD storage ($\$$400), so \thiswork with 1-3 CSDs shows lower GFLOPS/$\$$ than the baseline.
However, when using more than four SSDs, the speedup over the baseline makes \thiswork more efficient.
When scaling the number of SmartSSDs, the GFLOPS/$\$$ keeps increasing, showing the efficiency of \thiswork.

\subsection{Application Case Study: Fine-tuning}
\label{sec:acc}

Fine-tuning is a representative application where \thiswork can be applied because it requires loading the entire model and the optimizer states but has a relatively small training time. 
Thus, we demonstrate fine-tuning results with pre-trained LLMs on various datasets using \thiswork. %


For experiments,
publicly available pre-trained weights were used to conduct fine-tuning tasks.
We obtained pre-trained weights of BERT-345M from Megatron-LM GitHub~\cite{megatron-lm}, and GPT-2 (774M and 1.6B) from the huggingface hub~\cite{huggingface_hub}.
\cref{tab:bert_finetuning} shows the development set results of four datasets included in the GLUE benchmark~\cite{glue}.
For BERT-345M, we trained the model with 10 epochs for MNLI and 12 epochs for QQP following~\cite{megatron-lm}.
The model was trained with 3 epochs for SST-2 and QNLI to follow hyperparameters of~\cite{bert}.
We fixed the batch size as 4, which is the same batch size as the default experimental setup in \cref{sec:env}.
For GPT models, the model was finetuned for 3 epochs.
To use the mixed precision training and pre-trained weights from the huggingface, the accuracy was measured with the autocast feature, which performs FP32 operations for the numerically unstable operations (e.g., softmax).
The results show that \thiswork can stably achieve fine-tuning accuracy for various datasets. 
\smartupdate is algorithmically identical to the baseline training, so the accuracy is exactly the same as the baseline. 
\SmartTopK adopts lossy compression, but achieves comparable accuracy in all datasets.

\begin{table}
    \centering
    \caption{
        Finetuning accuracy and speedup comparison.
    }
    \label{tab:bert_finetuning}
    \resizebox{\columnwidth}{!}
    {
    \setlength{\tabcolsep}{3pt}
    \begin{tabular}{ccccccc}
    \toprule
         \multirowcell{2}{Model} & \multirowcell{2}{Method}& \multirowcell{1}{Speedup} &  \multicolumn{4}{c}{Accuracy (\%)}\\
         \cmidrule(lr){4-7}
         & & \multicolumn{1}{c}{(\#SSDs=6)} & \multicolumn{1}{c}{MNLI m/mm}  & \multicolumn{1}{c}{QQP} & \multicolumn{1}{c}{SST-2} & \multicolumn{1}{c}{QNLI} \\
         \midrule
          \multirowcell{6}{BERT-0.34B}& Baseline & 1$\times$ & 89.60/89.46 & 91.95 & 93.98 & 94.23 \\
          & SU+O & 1.10$\times$ & 89.60/89.46 & 91.95 & 93.98 & 94.23 \\
          & SU+O+C (10\%)& 1.23$\times$ & 89.42/89.55 & 91.69 & 94.95 & 94.16 \\
          & SU+O+C ( 5\%)& 1.34$\times$ & 89.24/89.38 & \rev{91.90} & 95.41 & 93.96 \\
          & SU+O+C ( 2\%) & 1.38$\times$ & 89.50/89.39 & 91.80 & 95.41 & 94.30 \\
          & SU+O+C ( 1\%) & 1.40$\times$ & 89.61/89.33 & 91.75 & 95.53 & 94.03 \\
        \midrule
          \multirowcell{6}{GPT2-0.77B}& Baseline & 1$\times$ &85.95/86.60 & 90.96 & 94.27 &91.60  \\
          & SU+O & 1.11$\times$ &85.95/86.60 & 90.96 & 94.27 &91.60  \\
          & SU+O+C (10\%)& 1.23$\times$ & 85.71/86.10 & 90.71 & 94.27 & 91.21 \\
          & SU+O+C ( 5\%)& 1.29$\times$ & 85.31/85.87 & 90.38 & 94.38 & 91.30 \\
          & SU+O+C ( 2\%)& 1.35$\times$ &85.05/85.25  &90.00  & 94.27 & 90.83 \\
          & SU+O+C ( 1\%)& 1.44$\times$ & 84.57/85.39 & 89.77 & 94.27 & 90.66 \\
        \midrule
          \multirowcell{6}{GPT2-1.6B}& Baseline & 1$\times$ &87.32/87.00 & 91.45 &95.18  &92.22  \\
          & SU+O & 1.29$\times$ &87.32/87.00 & 91.45 &95.18  &92.22  \\
          & SU+O+C (10\%) & 1.45$\times$ & 86.71/87.23 & 91.15 & 94.84 & 91.74 \\
          & SU+O+C ( 5\%)& 1.54$\times$ & 86.88/87.16 & 91.04 & 94.38 & \rev{91.65} \\
          & SU+O+C ( 2\%)& 1.54$ \times$ & 86.67/86.71 &90.83  & 94.61 & 91.62 \\
          & SU+O+C ( 1\%)& 1.53$\times$ & 86.58/86.69 & 90.56 & 94.72 & 91.45 \\
        
        \bottomrule
        \end{tabular}
    }
    
\end{table}

\subsection{Sensitivity of Compression Ratio}

\begin{figure}[t]
    \centering
    \includegraphics[width=\columnwidth]{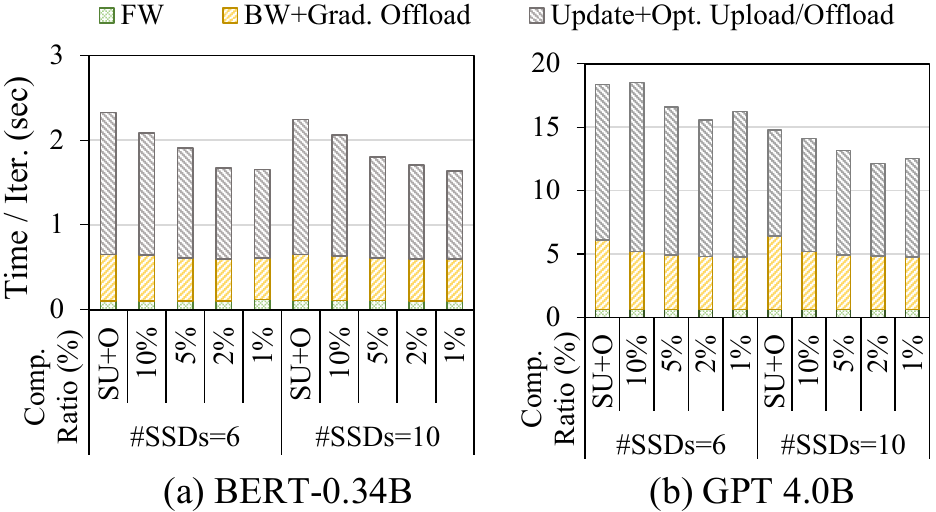}       
    \caption{
        Training time sensitivity on Top-K compression ratio of \thiswork. 
    } 
    \label{fig:comp_sensi}
\end{figure}

As a representative gradient compression, \SmartTopK uses a magnitude-based one.
We used 2\% compression (top 1\% selection, an index-value pair per element) as a default because it is the usual practice in various works~\cite{dgc, gradzip}.
To further investigate how \thiswork is sensitive to compression ratio, 
we experimented on various compression ratios from 1$\sim$10\% in \cref{tab:bert_finetuning} and \cref{fig:comp_sensi}.
There exists some trade-off between speedup and model quality.
However, it does not significantly harm model quality even with much more compression than the default, but the speedup almost gradually increases.



\section{Discussion}

\subsection{An Alternative Scenario: Multi-GPU Congested Topology} 
\begin{figure}[t]
    \centering
        \includegraphics[width=\columnwidth]{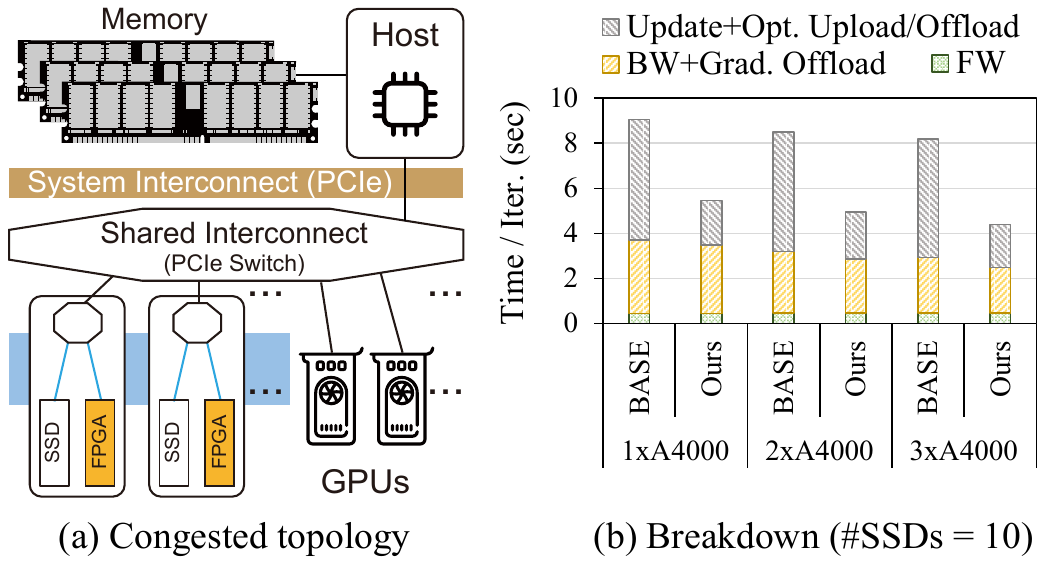}
    \caption{
        (a) An example of multi-GPU congested topology. (b) Training time breakdown of \thiswork and the baseline on such environment with 1$\sim$3 GPUs (GPT-2 1.16B).
    } 
    \label{fig:multi_gpu}
\end{figure}

When PCIe lanes of a system are more limited, GPUs and storages could be installed in the same PCIe expansion sharing the same PCIe switch, so the peer-to-peer communication topology changes.
To test such a scenario, we additionally equipped a congested topology with up to three GPUs, as illustrated in \cref{fig:multi_gpu}(a).
Due to the chassis limit of PCIe expansion, we inserted single-slot RTX A4000 GPUs into the expansion.
\cref{fig:multi_gpu}(b) shows the training time and speedup of \thiswork over the baseline when using 1$\sim$3 A4000 GPUs.
We adopted tensor parallelism for the multi-GPU strategy because it is widely selected in a single server setting~\cite{megatron-lm}. 
After gradients are calculated, each GPU identifies CSDs that own the corresponding parameters and performs updates with \thiswork. 

Using tensor parallelism slightly reduces `FW' and `BW' time when utilizing more GPUs.
However, in this alternative scenario, the remotely placed GPUs incur some extra traffic of model and activation transfer, which has to share the same PCIe interconnect with the CSDs. 
This causes some overhead to the `BW+Grad. Offload' phase compared to the default topology, while not greatly affecting the time of `Update+Opt. Upload/Offload' phase where the GPUs are idle.
%
Because of this, the observed speedup is smaller than the default setup, which indicates that the performance is affected by how PCIe topologies are structured.
However, \thiswork still provides 1.66$\times$$\sim$1.86$\times$ speedup with ten CSDs, which shows that it could be extended to a more PCIe lane-limited environment with multiple GPUs.

\subsection{Applying \thiswork to Model Compression}
\label{sec:dis_comp}
In this work, we have used fine-tuning to demonstrate the usefulness of \thiswork.
Various types of model compression methods can be usecases of \thiswork, such as quantization~\cite{pact, hawq}, pruning~\cite{prune, deepcompression}, or low-rank decomposition~\cite{pufferfish} because they require some fine-tuning in compressed form to recover accuracy drop from compression.

Interestingly, those applications are expected to bring even more speedup to \thiswork.
As discussed in \cref{sec:smartssd_topk}, the current bottleneck of \thiswork is on the upstream model transfer from the CSD to the host. 
When \thiswork is used for model compression, \thiswork can perform compression and upload the compressed model, further reducing the bottleneck.

However, achieving further speedup from this would bring some non-trivial issues.
For example, quantization often performs backpropagation with straight-through estimator (STE)~\cite{ste} with variable (floating point) quantization intervals per layer.
This means that the training GPUs use floating point models instead of integers, rather counter-intuitively.
To address the issue, the CSDs would have to derive the per-layer quantization intervals, convert the models to integers, and send the parameters upstream along with the interval values.
Then, the GPUs convert the integer weights into floating points using the intervals, so that it can perform STE.
For pruning and low-rank decomposition, similar issues can be found.
This indicates that the CSD now has to handle a more computationally complex job of compression using a lightweight FPGA, and an efficient GPU kernel for decompression has to be developed to not introduce additional bottlenecks. 
We leave this as a future work, and foresee more interesting approaches for model compression on top of \thiswork.

\subsection{Storage Expansion and Pooling}
\label{sec:expansion}
\thiswork utilizes a storage expansion system using PCIe switches to accommodate multiple CSDs. 
As modern workloads such as LLMs, recommendation, big data analytics~\cite{biscuit}, or bioinformatics demand more memory and storage capacity, new proposals are being made on multiple servers sharing storages~\cite{falcon} and memories~\cite{cxl}.
Especially when a large capacity is in need, often the choice is to equip such switches to multiply available slots for memory or storage.
These are often pooled among multiple servers to adapt to dynamic capacity requirements over time, and some provide direct GPU-initiated NVMe accesses~\cite{bam}.
We believe \thiswork is a great fit for such a movement.
Expansion using switches increases the number of physical slots for storage devices, but not the raw link bandwidth to the host.
When more devices are used for more capacity, then the link bandwidth bottleneck will only get severer.
Using techniques such as \thiswork with CSDs, more capacity means more internal bandwidth and computational capability to utilize them.
Along with such evolving system architecture toward sharing more resources, these types of solutions are expected to prevail.

\section{Related Work}

\subsection{Near-Data Processing} 
The idea of near-data processing has been studied, ranging from DRAM, SRAM, NVRAM to storage.
The early research on near-data processing was led by integrating DRAM with logic around 90's~\cite{execube, yukon, flexram, diva, iram, smartmem}.
Later, with the surge of 3D stacked memories~\cite{hmc}, several ideas were suggested to utilize the logic dies stacked with memory dies for various applications~\cite{tesseract, pei, graphp, graphq, graphpim, toppim, neurocube, tetris, dracc, xnorpop, googleworkloads}.
Recently, driven by the discontinuity of the 3D stacked memory, both academia and industry started turning back to the DDR variants, sometimes by modifying the die-internal circuitry~\cite{rowclone, ambit, bufcmp, drisa, mvid, hbmpim, aim, upmem, newton}, using buffer chips on DIMMs~\cite{chameleon, tensordimm, axdimm, nda}, or both~\cite{gradpim, trim}. 
While the principles are similar, these are backed by commercial products~\cite{hbmpim, upmem, aim}.

Such movements have also been made on storage devices.
Starting from ideas of separate accelerators near the IO subsystem~\cite{netezza, exadata}, earlier ideas were to utilize the embedded cores for computational approaches, led by database or big data workloads~\cite{smssdwisconsin, smssducsc, activeflash}. 
There were many follow-up approaches targeting various issues~\cite{recssd, summarizer, graphssd, iceclave, bluedbm, mithrilog}.
However, the embedded cores often turned out to be low in computational power for many applications~\cite{activeflash, activedisk, insider}, 
and more approaches started having dedicated accelerators.
ASICs were used for genome sequence analysis~\cite{genstore}, private information retrieval~\cite{inspire}, or ML queries~\cite{deepstore}. 
Some approaches employ FPGAs to enhance flexibility targeting similar issues~\cite{grafboost, insider, rmssd, extrav, metalfs, secNDP}.
It is worth noting GradPIM~\cite{gradpim} and OptimStore~\cite{optimstore} as they both target DNN training similar to this work. 
However, they assume dedicated memory/storage die, making those solutions less practical. Moreover, they were only implemented on simulators and do not consider real system integration issues.

Recently, several commercial products came out, equipped with an FPGA accelerator in the SSD package~\cite{eidetic, scaleflux, smssdsummit, polardb} for near-storage processing.
SmartSSD~\cite{smssdsummit} is a representative one, 
which was especially used for DB workloads~\cite{smssdcal} and sorting~\cite{nascent, nascent2}.
A similar platform was deployed in a commercial cloud for DB scans~\cite{polardb}. 
There are some works~\cite{near1,deep,smartsage} utilizing CSD for near-storage processing to accelerate training neural networks, but they mainly focus on preprocessing training data or embedding tables to reduce storage overhead. 
To the best of our knowledge, \thiswork is the first work using multiple CSDs to mitigate system interconnect bottlenecks in DNN training.
Our work builds on SmartSSD but is not limited to certain products.








\subsection{DNN Training Acceleration}

As deep learning models grow and training time becomes longer~\cite{efficientNet, bert, turing-nlg, gpt2}, many previous works aimed to reduce training time. 
Distributed training is one attractive way of using multiple workers. 
Data parallelism~\cite{trainingImagenetInHour,largeScaleDistributed, scaleOutLargeMinibatchSGD, elasticAveragingSGD} replicates the entire model, and each worker holds it to process training batch independently. 
However, when the model size exceeds the worker memory limit, the model needs to be split across workers, and model parallelism~\cite{krizhevsky2014one, chen2016training, chen2018efficient} enables training such large models.
Pipelining approach~\cite{gpipe, pipedream, pipemare, pipebd} has been proposed to address a low utilization of workers in model parallelism.
Each worker concurrently participates by dividing a minibatch into smaller microbatches and overlapping them.

In this distributed training, communication among servers becomes a bottleneck. 
Gradient compression is a popular approach to reduce such communication overhead. 
It effectively reduces training time while achieving comparable accuracy, even in large models~\cite{dalle, optimus_cc}. 
It can largely be categorized into two groups:
gradient sparsification~\cite{bernstein2018signsgd, dgc, scalecom} and low-rank decomposition~\cite{powersgd, gradzip}. 
The former sends some portion of gradients, usually using the magnitude criteria. 
Its convergence was proven in previous works~\cite{topkconverge, topkunderstand}, and it showed comparable accuracy. 
Low-rank decomposition factorizes a gradient matrix into two small low-rank matrices to reduce the communication volume.
\cite{powersgd} used the power iteration method to reduce the decomposition overhead. 
Both approaches use error compensation, which memorizes errors from the compression and adds them to gradients at the next step before compression. 
~\cite{1bit-adam} found that error compensation does not apply to the Adam optimizer due to its nonlinearity. 
Therefore, it preconditioned the variance term after a warm-up period to make it equivalent to momentum SGD.

\section{Conclusion}
We propose \thiswork, a novel CSD-based framework to accelerate storage-offloaded LLM training.
By utilizing the proposed FPGA accelerators in each CSD, the transfer bottleneck from LLM training is efficiently addressed.
We further provide transfer optimizations and CSD-assisted gradient compression to boost the speedup of the proposed system.
\thiswork has been fully integrated into PyTorch only with off-the-shelf products, making itself a ready-to-use solution. 
Experimental results show that \thiswork achieves a significant speedup, and shows good scalability to the increased number of CSDs attached to the system. 
Our implementation is available at \url{https://github.com/AIS-SNU/smart-infinity}.


\section*{Acknowledgements}
This work was supported by Samsung Electronics Co., Ltd (IO221213-04119-01), 
the National Research Foundation of Korea (NRF) grant funded by the Korea government (MSIT) (2022R1C1C1011307, 2022R1C1C1008131), 
and Institute of Information \& communications Technology Planning \& Evaluation (IITP) under the artificial intelligence semiconductor support program to nurture the best talents (IITP-2023-RS-2023-00256081) grant funded by the Korea government (MSIT). 
The experimental environment was provided by Samsung Memory Research Center (SMRC). 
Hongsun Jang, Jaewon Jung, Youngsok Kim, and Jinho Lee were partly supported by the BK21 FOUR (Fostering Outstanding Universities for Research) funded by the Ministry of Education (MOE) and the National Research Foundation (NRF) of Korea. 



\bibliographystyle{IEEEtranS}
\bibliography{refs}

\end{document}